\documentclass[11pt]{article}
\usepackage{hyperref}
\usepackage{latexsym}
\usepackage{amsmath}
\usepackage{amssymb}
\usepackage{amsfonts}

\def\01{\{0,1\}}

\newcommand{\eps}{\varepsilon}

\newcommand{\Pos}{\mbox{\rm Pos}}
\newcommand{\ket}[1]{|#1\rangle}
\newcommand{\bra}[1]{\langle#1|}
\newcommand{\ketbra}[2]{|#1\rangle\langle#2|}
\newcommand{\norm}[1]{{\| #1 \|}}
\newcommand{\inp}[2]{\langle{#1}|{#2}\rangle} 
\newcommand{\Tr}{\mbox{\rm Tr}}
\newcommand{\dono}{{\mathcal{?}}}

\newtheorem{theorem}{Theorem}
\newtheorem{lemma}{Lemma}

\newtheorem{corollary}{Corollary}

\newtheorem{claim}{Claim}

\newcommand{\qedsymb}{\hfill{\rule{2mm}{2mm}}}
\newenvironment{proof}[1][]{\begin{trivlist}
\item[\hspace{\labelsep}{\bf\noindent Proof#1:\/}] }{\qedsymb\end{trivlist}}

\begin{document}

\title{Bounded-Error Quantum State Identification and\\ Exponential Separations in Communication Complexity}
\author{Dmitry Gavinsky\thanks{University of Calgary. Supported in part by Canada's NSERC.}
 \and Julia Kempe\thanks{CNRS \&\ LRI, Univ.~de Paris-Sud, Orsay. Supported in part by ACI S\'ecurit\'e Informatique
SI/03 511 and ACI-Cryptologie CR/02 20040 grants of the French Research Ministry and the EU fifth framework project
RESQ, IST-2001-37559. Hospitality of the MSRI, Berkeley, where part of this work was done, is gratefully acknowledged.}
 \and Oded Regev\thanks{Department of Computer Science, Tel-Aviv University, Tel-Aviv 69978, Israel. Supported by an
 Alon Fellowship, by the Binational Science Foundation, and by the Israel Science Foundation.}
 \and Ronald de Wolf\thanks{CWI, Amsterdam. Supported by a Veni grant
from the Netherlands Organization for Scientific Research (NWO) and by the EU fifth framework project
 RESQ, IST-2001-37559.}}
\date{}
\maketitle
\thispagestyle{empty}

\begin{abstract}
We consider the problem of bounded-error quantum state identification: given either state $\alpha_0$ or state $\alpha_1$, we are
required to output `$0$', `$1$' or `$\dono$' (``don't know"), such that conditioned on outputting `$0$' or `$1$', our
guess is correct with high probability. The goal is to maximize the probability of not outputting `$\dono$'. We prove a
direct product theorem: if we're given two such problems, with optimal probabilities $a$ and $b$, respectively, and the
states in the first problem are pure, then the optimal probability for the joint bounded-error state identification
problem is $O(ab)$. Our proof is based on semidefinite programming duality and may be of wider interest.

Using this result, we present two exponential separations in the simultaneous message passing model of communication
complexity. Both are shown in the strongest possible sense. First, we describe a relation that can be computed with
$O(\log n)$ classical bits of communication in the presence of shared randomness, but needs $\Omega(n^{1/3})$
communication if the parties don't share randomness, even if communication is quantum. This shows the optimality of
Yao's recent exponential simulation of shared-randomness protocols by quantum protocols without shared randomness.
Second, we describe a relation that can be computed with $O(\log n)$ classical bits of communication in the presence of
shared entanglement, but needs $\Omega((n/\log n)^{1/3})$ communication if the parties share randomness but no
entanglement, even if communication is quantum. This is the first example in communication complexity of a situation
where entanglement buys you much more than quantum communication does.
\end{abstract}

\newpage
\setcounter{page}{1}

\section{Introduction}

\subsection{Bounded-error quantum state identification}

Suppose we are given one of two mixed quantum states, $\alpha_0$ or
$\alpha_1$, each with probability 1/2.  We know what $\alpha_0$ and $\alpha_1$ are.
Our goal is to identify which one we are given.
It is well known that we can output the correct answer (0 or~1) with probability
$1/2+\norm{\alpha_0-\alpha_1}_{tr}/2$, where $\norm{\cdot}_{tr}$ is
the trace norm (the sum of the singular values, divided by 2).
This is optimal. In particular, if $\alpha_0$
and $\alpha_1$ are very close in trace norm, the best measurement will
do little better than a fair coin flip. In some situations, however,
we cannot afford to output the wrong answer with such high probability, and
would rather settle for a measurement that sometimes claims ignorance,
but that is usually correct in the case where it does give an output.

To illustrate this, suppose the states involved are the following pure states:
\begin{center}
$\ket{\alpha_0}=\sqrt{a}\ket{0}+\sqrt{1-a}\ket{2}$\\
$\ket{\alpha_1}=\sqrt{a}\ket{1}+\sqrt{1-a}\ket{2}$
\end{center}
If we cannot afford to make a mistake at all, it is clear what measurement we should
apply: measure in the computational basis, and if the outcome is 0 the state must have been
$\alpha_0$; if the outcome is 1 the state must have been $\alpha_1$; if the outcome is 2
we claim ignorance. Note that the probability of getting an answer (0 or 1)
for the identification problem is now only $a$. We have thus increased our confidence
in the answer, at the expense of decreasing the probability of getting an answer at all.
Now consider a slightly more ``fudged'' example, for some small $\eps$:
\begin{center}
$\ket{\alpha_0}=\sqrt{(1-\eps)a}\ket{0}+\sqrt{\eps a}\ket{1}+\sqrt{1-a}\ket{2}$\\
$\ket{\alpha_1}=\sqrt{\eps a}\ket{0}+\sqrt{(1-\eps)a}\ket{1}+\sqrt{1-a}\ket{2}$
\end{center}
If we apply the same procedure as before, we have now a small probability of error:
on both states our measurement outputs a guess (0 or 1) with probability $a$,
and \emph{if} we output a guess, then that guess is wrong with probability only $\eps$.
If $\eps$ is sufficiently small, this may still be acceptable for many applications.

More generally, let $A$ be some classical random variable, and $B$ be another random variable whose range includes the
special symbol  `$\dono$'. We call $B$ an \emph{$(a,\eps)$-predictor} for $A$ if $\Pr[B\neq \dono]\geq a$ and
$\Pr[A=B\mid B\neq \dono]\geq 1-\eps$. For example, the above measurement applied to state $\alpha_X$ where $X$ is a
random bit, gives us an $(a,\eps)$-predictor for $X$ if we interpret output 2 as `$\dono$'. Motivated by the above
examples---and by our applications in later sections---we define the bounded-error state identification problem:
\begin{quote}
Given a register containing $\alpha_X$, with $X$ a uniformly random bit, and an $\eps >0$, what is the maximal $a$ for
which there exists a quantum measurement on the register whose outcome is an $(a,\eps)$-predictor for $X$?
\end{quote}
We use $D_\eps(\alpha_0,\alpha_1)$ to denote the maximal value $a$. We stress again that the error probability is a
\emph{conditional} probability, conditioned on actually outputting a guess for the bit (0 or~1). Unlike the
straightforward distinguishing problem, where the optimal success probability is determined by the trace distance
$\norm{\alpha_0-\alpha_1}_{tr}$, we do not know of any simple metric on density matrices that determines the value
$D_\eps(\alpha_0,\alpha_1)$. However, as was also noted by~Eldar~\cite{eldar:sdp}, one can easily express quantities
like this as the optimal value of a semidefinite program, as we will do in Section~\ref{secsdpformulation}.

Now suppose we are given another identification problem in a second register, quantum state $\beta_Y$ for a random bit
$Y$, and suppose $b=D_\eps(\beta_0,\beta_1)$ is the largest value for which we can obtain a $(b,\eps)$-predictor for
$Y$. We now want to determine the optimal probability with which we can identify (again with error at most $\eps$ or
something related) \emph{both} states simultaneously. That is, what is the maximal probability
$p=D_\eps(\alpha_0\otimes\beta_0,\alpha_0\otimes\beta_1,\alpha_1\otimes\beta_0,\alpha_1\otimes\beta_1)$ such that a
joint measurement on $\alpha_X\otimes\beta_Y$ gives us a $(p,\eps)$-predictor for $XY$? Since the two registers are
completely independent, it seems there is nothing much better we can do except applying the optimal measurement for
both registers separately.\footnote{This actually gives slightly worse error $2\eps-\eps^2$ for the prediction of $XY$,
so potentially it could be that $p \ll ab$.} Thus our intuition suggests that $p\leq ab$, or at least $p\leq O(ab)$.
This problem has a flavor similar to ``direct product theorems'' in computational complexity theory, where one is
usually interested in $k\geq 2$ independent instances of some computational problem, and the aim is
to show that the overall success probability of some algorithm for the $k$-fold problem is close to
the product of the $k$ individual success probabilities. Another problem with a similar
flavor is the notoriously hard quantum information theory issue of multiplicativity of norms of superoperators under
tensor product~\cite{king&ruskai:multi}.

Proving our intuition actually turned out to be quite a hard problem, and we indicate some reasons why in
Section~\ref{secwhyhard}. In an earlier preprint~\cite{gkw:qsm} we were only able to prove it for $\eps=0$, which was
then used by us in~\cite{gkw:qsm} and~\cite{gavinsky:note} to obtain various zero-error separations in communication
complexity. The present paper supersedes all of these unpublished results and gives in
Section~\ref{secdirectproduct} the first proof of the $p\leq O(ab)$ bound for the case where at least one of the two
sides is pure (i.e., $\alpha_0$ and $\alpha_1$ are both pure, or $\beta_0$ and $\beta_1$ are both pure). More
precisely, we show
\begin{equation}\label{eq:main_stat}
D_{\eps/2}(\alpha_0\otimes\beta_0,\alpha_0\otimes\beta_1,\alpha_1\otimes\beta_0,\alpha_1\otimes\beta_1) \leq
O(D_\eps(\alpha_0,\alpha_1)\cdot D_\eps(\beta_0,\beta_1)).
\end{equation}
Notice that because of the $\eps/2$ on the left hand side, this bound is slightly weaker than what we have promised; as
we indicate in Section~\ref{secwhyhard}, this modification is (somewhat surprisingly) necessary. Our proof relies
heavily on a semidefinite programming formulation for the quantities involved and on an analysis of their duals.

\subsection{Exponential separations in communication complexity}

Apart from being an interesting information theoretic problem in its own right, the bounded-error state identification
problem and our direct product theorem have interesting applications. We give two new exponential separations, both in
the simultaneous message passing (SMP) model of communication complexity. The area of communication complexity deals
with the amount of communication required for solving computational problems with distributed input. This area is
interesting for its own sake, but also has many applications to lower bounds on circuit size, data structures, etc. The
\emph{simultaneous message passing} (SMP) model involves three parties: Alice, Bob, and a referee. Alice gets input
$x$, Bob gets input $y$. They each send one message to the referee, to enable him to compute something depending on
both $x$ and $y$, such as a Boolean function or some relational property. The \emph{cost} or \emph{complexity} of a
communication protocol is the length of the total communication for a worst-case input, and the complexity of a problem
is the cost of the best protocol that solves our problem with small error probability.

The SMP model is arguably the weakest setting of communication complexity that is still interesting. Even this simple
setting is not well understood. In the case of deterministic protocols, the optimal communication is determined by the
number of distinct rows (and columns) in the communication matrix, which is a simple property. However, as soon as we
add randomization to the model things become much more complicated.  For one, we can choose to either add \emph{shared}
(a.k.a.~public) or \emph{private} randomness. In other communication models this difference affects the optimal
communication by at most an additive $O(\log n)$~\cite{newman:random}, but in the SMP model the difference can be huge.
For example, the equality function for $n$-bit strings requires about $\sqrt{n}$ bits of communication if the parties
have only private randomness \cite{ambainis:3computer,newman&szegedy:1round,babai&kimmel:simultaneous}, but only
constant communication with shared randomness! No simple characterization of SMP communication complexity with either
private or shared randomness is known.\footnote{Kremer et~al.~\cite{knr:rand1round} claimed a characterization of
shared-randomness complexity as the largest of the two one-way complexities, but Bar-Yossef et
al.~\cite[Section~4]{bjks:itcc} exhibited a function where their characterization fails.}

The situation becomes more complicated still when we throw in \emph{quantum} communication. Buhrman
et~al.~\cite{bcww:fp} exhibited a quantum protocol for the equality function with $O(\log n)$ qubits of communication.
This is exponentially better than classical private-randomness protocols, but slightly worse than shared-randomness
protocols. Roughly speaking, their quantum fingerprinting technique may be viewed as replacing the shared randomness by
a quantum superposition.

\subsubsection{Shared randomness beats quantum communication}

The fingerprinting idea of~\cite{bcww:fp} was generalized by Yao~\cite{yao:qfp}, who showed that every classical shared-randomness
protocol with $c$-bit messages for a Boolean function can be simulated by a quantum fingerprinting protocol that uses
$O(2^{4c}\log n)$ qubits of communication. This has since been improved to $O(2^{2c}\log n)$
qubits~\cite{gkw:qsm,golinsky&sen:qfp}. In particular, every $O(1)$-bit shared-randomness protocol can be simulated by
an $O(\log n)$-qubit quantum protocol. Again, quantum superposition replaces shared randomness in this construction.

This raises the question whether something similar always holds in the SMP model: can every classical shared-randomness
protocol be efficiently simulated by some protocol that sends qubits but shares neither randomness nor entanglement?
Since the appearance of Yao's paper, quite a number of people have tried to address this. Our first result, presented
in Section~\ref{secappl1}, gives a negative answer to this question. Suppose Alice receives inputs $x,s\in\01^n$ with
the property that $s$ has Hamming weight $n/2$ and Bob receives input $y\in\01^n$. The referee should output, with
probability at least $1-\eps$, a triple $(i,x_i,y_i)$ for an $i$ satisfying $s_i=1$. We prove that protocols where
Alice and Bob share randomness can solve this task with $O(\log n)$ classical bits of communication, while every
bounded-error quantum protocol without shared randomness needs $\Omega(n^{1/3})$ qubits of communication. The quantum
lower bound relies crucially on our direct product theorem for bounded-error state identification. This shows for the
first time that the resource of shared randomness cannot be efficiently traded for quantum communication.

Yao's exponential simulation can be made to work for relations as well,
and our quantum lower bound shows that it is essentially optimal,
since the required quantum communication is exponentially larger
than the classical shared-randomness complexity for our relational problem.
We expect a similar gap to hold for (promise) Boolean functions as well.
Our separation complements a separation in the other direction: Bar-Yossef et al.~\cite{bjk:q1way} exhibited a relation
where quantum SMP protocols are exponentially \emph{more} efficient than classical SMP protocols even with shared
randomness (also in their case it is open whether there is a similar gap for a Boolean function). Accordingly, the
quantum SMP model is incomparable with the classical shared-randomness SMP model.

\subsubsection{Shared entanglement beats quantum communication with shared randomness}

The second application of our state identification result is again in the SMP model.
While the previous application separated classical protocols with
shared randomness from quantum protocols without shared randomness,
this one separates classical protocols with \emph{entanglement}
(EPR-pairs, 2-qubit states of the form $\frac{1}{2}(\ket{00}+\ket{11}))$
from quantum protocols with shared randomness.

The additional power that prior entanglement gives is one of the most fundamental questions in quantum communication
complexity. This additional power is not well understood. We basically know two ways in which entanglement can help: it
can be used for teleportation (where one EPR-pair and two classical bits of communication replace one qubit of
communication) and it can be used for shared randomness (if Alice and Bob each measure their side of their shared
EPR-pair in the computational basis, they get the same random bit). Neither saves very much communication, and it has
in fact been conjectured for the standard two-party one-round and many-round protocols that the model of classical
communication with entanglement~\cite{cleve&buhrman:subs} and the model of quantum communication 
without entanglement~\cite{yao:qcircuit} are essentially equivalent.

Our second separation shows that the situation is very different in the simultaneous message passing model. We show that
the qubit-communication model cannot efficiently simulate the entanglement model. In Section~\ref{secappl2} we exhibit
a relational problem, inspired by the problem of Bar-Yossef et al.~mentioned above, that can be solved with $\log n$
EPR-pairs shared between Alice and Bob and $O(\log n)$ classical bits of communication. In contrast, if only shared
randomness is available instead of entanglement, every bounded-error SMP protocol needs $\Omega((n/\log n)^{1/3})$
quantum bits of communication. Again, our direct product theorem is crucial for proving the quantum lower bound. This
is the first example of a communication problem where entanglement is much more useful than quantum communication.

\section{Preliminaries}

\subsection{Quantum computing}

The essentials needed for this paper are quantum states and their measurement.  First, an $m$-qubit \emph{pure state}
is a superposition $\ket{\phi}=\sum_{z\in\01^m}\alpha_z\ket{z}$ over all classical $m$-bit states. The $\alpha_z$'s are
complex numbers called \emph{amplitudes}, and $\sum_z|\alpha_z|^2=1$. Hence a pure state $\ket{\phi}$ is a unit vector
in $\mathbb{C}^{2^m}$. Its complex conjugate (a row vector with entries conjugated) is denoted $\bra{\phi}$. The inner
product between $\ket{\phi}$ and $\ket{\psi}=\sum_z\beta_z\ket{z}$ is the dot product
$\bra{\phi}\cdot\ket{\psi}=\inp{\phi}{\psi}=\sum_z\alpha_z^*\beta_z$. The \emph{norm} of a vector $v$ is
$\norm{v}=\sqrt{\inp{v}{v}}$. Second, a \emph{mixed state} $\rho=\sum_i p_i\ketbra{\phi_i}{\phi_i}$ corresponds to a
probability distribution over pure states, where $\ket{\phi_i}$ is given with probability $p_i$. A $k$-outcome
\emph{positive operator-valued measurement} (POVM) is given by $k$ positive semidefinite operators $E_1,\ldots,E_k$
with the property that $\sum_{i=1}^k E_i=I$. When this POVM is applied to a mixed state $\rho$, the probability of the
$i$-th outcome is given by the trace $\Tr[E_i\rho]$. We refer to Nielsen and Chuang~\cite{nielsen&chuang:qc} for more
details.

\subsection{Communication complexity}

We now give a somewhat informal description of the simultaneous message passing model discussed in our two
applications. For a more formal description, we refer to Kushilevitz and Nisan~\cite{kushilevitz&nisan:cc} for
classical communication complexity and to the surveys~\cite{klauck:qccsurvey,buhrman:qccsurvey,wolf:qccsurvey} for the
quantum variant. In the simultaneous message passing model, Alice receives input $x$, Bob receives input $y$, they each
send a message to a referee who should then output either $f(x,y)$ in the case of a functional problem, or an element
from some set $R(x,y)$ in the case of a relational problem. We use $R_\eps^{\parallel}(P)$,
$R_\eps^{\parallel,pub}(P)$, $R_\eps^{\parallel,ent}(P)$ to denote, respectively, the optimal communication complexity
of classical protocols that solve problem $P$ with worst-case error probability $\eps$, using, respectively, private
randomness, shared randomness between Alice and Bob, and shared entanglement between Alice and Bob (EPR pairs).
The number of shared coin flips or shared EPR-pairs is unlimited and does not count towards the communication cost of
the protocol. We use $Q_\eps^{\parallel}(P)$, $Q_\eps^{\parallel,pub}(P)$, $Q_\eps^{\parallel,ent}(P)$ for the variety
that allows quantum communication.

\subsection{The random access code argument}\label{secrac}

Here we will describe a slight extension of a quantum information theory argument due to Ashwin Nayak~\cite{nayak:qfa}
that we will apply several times in our communication complexity lower bounds.
We call this the ``random access code argument''.
In the following, we assume familiarity with basic classical information
theory~\cite{cover&thomas:infoth} and quantum information theory~\cite{nielsen&chuang:qc}.

\begin{lemma}\label{lem:RAC}[``Random Access Code Argument"]
Let $X=X_1\ldots X_n$ be a classical random variable of $n$ uniformly distributed bits. Suppose for each instantiation
$X=x$ we have a quantum state $M_x$ of $q$ qubits. Suppose also that for each $i\in[n]$ of our choice we can apply a
quantum measurement to $M_X$ whose outcome is a $(\lambda_i,\eps_i)$-predictor for $X_i$. Then
$$\sum_{i=1}^n\lambda_i(1-H(\eps_i))\leq q.$$
\end{lemma}

\noindent
Before giving the proof, notice the following special case: if we can predict each $X_i$ with bias $\eta_i$
(i.e., we have a $(1,1/2-\eta_i)$-predictor), then the above bound becomes
$$
\sum_{i=1}^n(1-H(1/2-\eta_i))\leq q.
$$
Since $1-H(1/2-\eta_i)=\Theta(\eta_i^2)$, the left hand side
is essentially the sum of squares of the $\eta_i$.

\begin{proof}
First, let $Y$ be a classical random variable corresponding to a uniformly distributed bit. Let $B$ be another random
variable that is a $(\lambda,\eps)$-predictor of $Y$. Using $H(Y \mid B, B \neq \dono) \leq H(\eps)$ 
and $\Pr[B \neq\dono] \geq \lambda$, we can upper bound the entropy of $Y$ given $B$:
\begin{eqnarray*}
H(Y\mid B) & = & \Pr[B = \dono] \cdot H(Y \mid B, B=\dono) + \Pr[B \neq \dono] \cdot H(Y \mid B, B \neq \dono)\\
 & \leq & (1-\Pr[B \neq \dono])\cdot 1 +  \Pr[B \neq \dono] \cdot H(\eps)\leq 1-\lambda(1-H(\eps)),
\end{eqnarray*}
and hence lower bound the mutual information between $Y$ and $B$:
$$
I(Y:B) = H(Y) - H(Y \mid B)\geq \lambda(1-H(\eps)).
$$
Now let $B_i$ be the outcome of the measurement corresponding to $i$ applied to $M_X$. 
We have
$$
S(X_i:M_X) \geq I(X_i:B_i) \geq \lambda_i(1-H(\eps_i))
$$
by Holevo's theorem~\cite{holevo} (the left hand side is equal to the Holevo $\chi$-quantity). 

Using~\cite[Theorem~11.8.5]{nielsen&chuang:qc} we have
$$
S(X:M_X)=S(X)+S(M_X)-S(X,M_X)=S(M_X)-\frac{1}{2^n}\sum_{x\in\01^n} S(M_x)\leq S(M_X)\leq q.
$$
Abbreviating $X_{1:i-1}=X_1\ldots X_{i-1}$, a chain rule for mutual information gives
$$
S(X:M_X)=\sum_{i=1}^n S(X_i:M_X\mid X_{1:i-1}).
$$
Using strong subadditivity and the fact that $S(X_i\mid X_{1:i-1})=S(X_i)$ we get
$$
S(X_i:M_X\mid X_{1:i-1})
=S(X_i\mid X_{1:i-1})-S(X_i\mid M_X X_{1:i-1})
\geq S(X_i)-S(X_i\mid M_X)
=S(X_i:M_X).
$$
Combining our inequalities gives the desired lower bound on $q$.
\end{proof}

\section{Bounded-error quantum state identification: Direct product}\label{secdirectproduct}

\subsection{Why this is delicate and non-trivial}\label{secwhyhard}

We briefly recall the 2-register state identification problem from the introduction. In the first register we are given
a quantum state $\alpha_X$, with $X$ a random bit, and the optimal probability with which we can get an
$\eps$-predictor for $X$ is $a$. In the second register we're given $\beta_Y$, with $Y$ a random bit, and the optimal
probability with which we can get an $\eps$-predictor for $Y$ is $b$. We now want to know the optimal probability $p$
with which a joint measurement on both registers can obtain an $\eps$-predictor for $XY$. As mentioned in the
introduction, intuition suggests that $p\leq O(ab)$. Before proceeding to prove a slightly weaker form of this
statement (namely the special case where $\alpha_0$ and $\alpha_1$ are pure), we will pause to sketch two variants of
the problem where the same intuition is provably \emph{false}, even for pure states! This points to the subtleness of
the state identification problem: seemingly small changes to the setup change everything.

First, suppose that instead of an $\eps$-predictor for $XY$ we want an
$\eps$-predictor for the parity $X\oplus Y$ of the two bits.
This might be slightly easier than getting both bits separately,
but intuition still suggests that because both registers are independent,
the best we can do is predict both registers separately and output their
parity if both measurements gave an answer. So we expect $p\leq O(ab)$.
However, this intuition is {\em false}.
Consider the following counterexample, with $\delta$ very small:
\begin{quote}
$\ket{\alpha_0}=\ket{\beta_0}=\ket{0}$\\
$\ket{\alpha_1}=\ket{\beta_1}=\sqrt{1-\delta^2}\ket{0}+\delta\ket{1}$
\end{quote}
It is not hard to convince oneself\footnote{A rigorous proof can be obtained from the SDP formulation of this problem.}
that for any fixed $\eps <1/2$, the optimal $a$ and $b$ are $\Theta(\delta^2)$, so our
intuition suggests $p\leq O(ab)=O(\delta^4)$ for the parity problem. However, if we apply the measurement with operator
$E_0$ that projects onto the state $\frac{1}{\sqrt{2+\delta^2}}(\delta \ket{00}-\ket{01}-\ket{10})$, $E_1=0$, and
$E_\dono=I-E_0$, then on the parity-0 inputs $\alpha_0\otimes\beta_0$ and $\alpha_1\otimes\beta_1$ the measurement
gives outcome 0 with probability roughly $\delta^2$, while on the parity-1 inputs it gives outcome 0 with probability
only about $\delta^6 \ll \eps \delta^2$. Thus, in this example $p$ is of the same order as $a$ and $b$ instead of their
product.

In our second example, we return to the original setting where we want to obtain a predictor for $XY$ (not their parity).
We consider the case where in the left hand side of Eq.~(\ref{eq:main_stat}) from the introduction we replace $\eps/2$ with a
slightly larger error parameter. Surprisingly, we show that in this case the bound $p\leq O(ab)$ is {\em false}. Choose
$\eps$ to be, say, $0.49$, and replace $\eps/2$ in the left hand side of (\ref{eq:main_stat}) with
something slightly larger, say, $0.251$.\footnote{With some effort, this example can be generalized to other values of $\eps$.}
To construct this example, we use the same states as in the previous example. For our choice of $\eps$, we still have
$a,b=\Theta(\delta^2)$. Now consider the measurement where operator $E_{00}$ projects onto the state
$\frac{1}{\sqrt{8/9+\delta^2}}(\delta\ket{00}-\frac{2}{3}\ket{01}-\frac{2}{3}\ket{10})$, $E_{01}=E_{10}=E_{11}=0$, and
$E_\dono=I-E_{00}$. Then on the state $\alpha_0\otimes\beta_0$ we get outcome 00 with probability roughly
$9\delta^2/8$, while on each of the other three states this probability is roughly $\delta^2/8$. Conditioned on outputting an
answer, our error probability is roughly $(3/8)/(9/8+3/8)=1/4$, so we obtain a $0.251$-predictor for $XY$. We see that again,
contrary to our intuition, $p$ is of the same order as $a$ and $b$.

Finally, to get a better feel for this problem and for why it is non-trivial, let us consider the classical case. This
is the special case of the problem in which all states involved are classical probability distributions. In other words
the density matrices $\alpha_0,\alpha_1$ are diagonal in the same basis and similarly for
$\beta_0,\beta_1$.\footnote{This is related to  optimal detector design, see e.g. \cite{vandenberghe&boyd:sdp},
Section~7.3.} In this case, one can give a characterization of the optimal measurement. Let $\alpha_0$ (resp.,
$\alpha_1$) correspond to some probability distribution on $n$ elements with probabilities $p_1,\ldots,p_n$ (resp.,
$q_1,\ldots,q_n$).  Assume without loss of generality that the $n$ elements are sorted by non-increasing order of
$\max\{p_i/(p_i+q_i), q_i/(p_i+q_i)\}$. For any $k \ge 1$, consider the measurement that maps the outcome $i$ for $1
\le i\le k$ to either $0$ if $p_i>q_i$ or $1$ otherwise, and maps any outcome $i>k$ to `$\dono$'. This means that for
each $i \le k$ we output the guess ($\alpha_0$ or $\alpha_1$) that is more likely, conditioned on $i$. Note that
$\max\{p_i/(p_i+q_i), q_i/(p_i+q_i)\}$ represents the probability that our guess is correct, given $i$. Then, for any
error parameter $\eps$, one can show that the best measurement is obtained by taking $k$ as large as possible while
still keeping the error probability of the resulting measurement below $\eps$.\footnote{To be precise, we should also
allow non-integer $k$ in the sense that when the outcome is $\lceil k \rceil$, one should output either $0$ or $1$
(depending on whether $p_{\lceil k \rceil} > q_{\lceil k \rceil}$) with probability $k-\lfloor k \rfloor$ and `$\dono$'
otherwise.}

Now assume we have probability distributions $\alpha_0,\alpha_1,\beta_0,\beta_1$ (equivalently, diagonal matrices) and
we want to predict $XY$ based on a sample from $\alpha_X \otimes \beta_Y$ (the tensor can be described
classically as one sample from $\alpha_X$ together with one independent sample from $\beta_Y$). The optimal measurement
in the two-register case can be obtained by a straightforward generalization of the measurement we have described in
the single register case.  As mentioned in the introduction, one might expect the optimal measurement to use the first
register to predict $X$ and the second register to predict $Y$ separately, i.e., to be a tensor product measurement. It
is perhaps somewhat surprising that this is {\em not} true in general, as can be seen using some simple examples. The
intuitive reason for this is that if a sample $(i,j)$ from $\alpha_X \otimes \beta_Y$ is such that $i$ gives a very
strong indication of (say) $\alpha_0$, then we might be willing to predict the state $\alpha_0 \otimes \beta_0$ even if
$j$ gives only a weak indication of $\beta_0$.

Nevertheless, the direct product theorem of Eq.~(\ref{eq:main_stat}) does hold in the classical case, even when we
replace $\eps/2$ with $\eps$. One proof of this is based on a similar approach to the one we will take in the quantum
case: first, formulate the problem in terms of linear programs (which are very similar to the semidefinite programs
that arise in the quantum case) and then bound the dual solution of the joint system. Bounding the dual solution is the
most demanding step technically, and amounts to solving some inequalities on real numbers. In the general quantum case,
this step involves some (rather nasty) matrix inequalities that seem quite difficult to solve. In the special case that
we consider below, these matrix inequalities turn out to have a sufficiently nice form to be analyzed.

\subsection{Proof of the direct product theorem}\label{secsdpformulation}

In this section we prove our main results about the 2-register quantum state identification problem. We use the
powerful technique of semidefinite programming duality. For details on semidefinite programming, see
e.g.~\cite{lovasz:sdp,vandenberghe&boyd:sdp}. We will actually prove two bounds. First, for the case where $\alpha_0$,
$\alpha_1$ are pure and $\beta_0$, $\beta_1$ are unrestricted, our Theorem~\ref{lemma:tensor} implies
\begin{equation}\label{eq:firsttensor}
D_{\eps/2}(\alpha_0\otimes\beta_0,\alpha_0\otimes\beta_1,\alpha_1\otimes\beta_0,\alpha_1\otimes\beta_1)\leq
O(D_{\eps}(\alpha_0,\alpha_1) \cdot D_{\eps}(\beta_0,\beta_1)).
\end{equation}
Second, if we allow all of $\alpha_0,\alpha_1,\beta_0,\beta_1$ to be mixed states then our Corollary~\ref{tensorcorollary} gives
$$
D_{\eps/2}(\alpha_0\otimes\beta_0,\alpha_0\otimes\beta_1,\alpha_1\otimes\beta_0,\alpha_1\otimes\beta_1)\leq
O(\|\alpha_0-\alpha_1\|_{tr}\cdot D_{\eps}(\beta_0,\beta_1)).
$$
The second bound will follow from the first by purifying the mixed states $\alpha_0$ and $\alpha_1$.

Let us first characterize $D_\eps(\alpha_0,\alpha_1)$. Recall that any measurement whose outcome is an
$(a,\eps)$-predictor outputs the correct answer with probability at least $1-\eps$ \emph{conditioned} on outputting a
guess (0 or 1, but not $\dono$). Denote the three measurement operators by $E_0$, $E_1$, $E_\dono$. Then we require
 \begin{equation}\label{eq:bound}
  \eps\geq\Pr[\mbox{wrong
guess}\mid\mbox{guess}]= \frac{\Pr[\mbox{wrong guess}]}{\Pr[\mbox{guess}]}=
\frac{\frac{1}{2}\Tr[E_0\alpha_1]+\frac{1}{2}\Tr[E_1\alpha_0]}{\Tr\left[(E_0+E_1)\alpha\right]},
 \end{equation}
where $\alpha=\frac{1}{2}(\alpha_0+\alpha_1)$ is the average state. To our knowledge there is no simple expression
for $D_\eps(\alpha_0,\alpha_1)$ in terms of $\alpha_0$ and $\alpha_1$. However, one can easily express it as a solution
to a semidefinite program (SDP). For fixed density matrices $\alpha_0$, $\alpha_1$ and fixed $\eps\in[0,1/2)$, the
optimal value $a=D_\eps(\alpha_0,\alpha_1)$ is given by the following SDP:
\begin{equation}\label{eq:primalsingle}
\begin{array}{ll}
\mbox{maximize}   & \Tr[(E_0+E_1)\alpha]\\
\mbox{subject to} & 0\preceq E_0, E_1,\\
                  & E_0+E_1 \preceq I,\\
                  & \frac{1}{2}\Tr[E_0\alpha_1]+\frac{1}{2}\Tr[E_1\alpha_0]\leq \eps\Tr[(E_0+E_1)\alpha].
\end{array}
\end{equation}
The first two constraints state that the operators $E_0,E_1$ together with a third operator $E_\dono=I-E_0-E_1$ form a
valid quantum measurement. The last constraint bounds the conditional error probability, as in Eq.~(\ref{eq:bound}). An
analogous SDP can be written for $b=D_\eps(\beta_0,\beta_1)$.

Similarly we can write the primal SDP that optimizes
$p=D_\eps(\alpha_0\otimes\beta_0,\alpha_0\otimes\beta_1,\alpha_1\otimes\beta_0,\alpha_1\otimes\beta_1)$:
\begin{equation}
\begin{array}{ll}
\mbox{maximize}   & \Tr[\left(E_{00}+E_{01}+E_{10}+E_{11}\right)\alpha\otimes\beta]\\
\mbox{subject to}  & 0\preceq E_{00}, E_{01}, E_{10}, E_{11},\\
                  & E_{00}+E_{01}+E_{10}+E_{11} \preceq I,\\
                  & \frac{1}{4}\Tr\left[\left(E_{01}+E_{10}+E_{11}\right)\alpha_0\otimes\beta_0+
                    \left(E_{00}+E_{10}+E_{11}\right)\alpha_0\otimes\beta_1+\right.\\
                  & \hspace{2em} \left.\left(E_{00}+E_{01}+E_{11}\right)\alpha_1\otimes\beta_0+
                    \left(E_{00}+E_{01}+E_{10}\right)\alpha_1\otimes\beta_1\right]\\
          & \hspace{2em} \leq \eps\Tr[\left(E_{00}+E_{01}+E_{10}+E_{11}\right)\alpha\otimes\beta].
\end{array}
\end{equation}
Here $\alpha \otimes \beta =
\frac{1}{4}(\alpha_0\otimes\beta_0+\alpha_0\otimes\beta_1+\alpha_1\otimes\beta_0+\alpha_1\otimes\beta_1)$ is the
average state.

\begin{theorem}\label{lemma:tensor}
Let $0 \leq \eps <\frac{1}{2}$ and $\alpha_0,\alpha_1$, $\beta_0,\beta_1$ be density matrices, where
$\alpha_0,\alpha_1$ correspond to pure states $\ket{\alpha_0},\ket{\alpha_1}$.
Let $b=D_{\eps}(\beta_0,\beta_1)$ and
$p=D_{\eps/2}(\alpha_0\otimes\beta_0,\alpha_0\otimes\beta_1,\alpha_1\otimes\beta_0,\alpha_1\otimes\beta_1)$.
Then
$$
p\leq 16(1-|\inp{\alpha_0}{\alpha_1}|^2)\cdot b.
$$
\end{theorem}

\noindent
 Since $\alpha_0$ and $\alpha_1$ are pure,
$a=D_\eps(\alpha_0,\alpha_1)\geq D_0(\alpha_0,\alpha_1) \geq\frac{1}{2}(1-|\inp{\alpha_0}{\alpha_1}|^2)$, where the
last inequality follows by considering the projective measurement on $\ket{\alpha_0}$ and $\ket{\alpha_0^\perp}$.
Hence this theorem implies Eq.~(\ref{eq:firsttensor}).

\begin{proof}
The idea behind our proof is the following. As we observed before, both $b$ and $p$ are the solution of an SDP and so
any feasible solution of the corresponding dual SDP yields an upper bound to $b$ resp.~$p$. We will show that a
feasible solution with value $d_b\geq b$ for the dual for $b$ can be used to construct a feasible solution with value
$16(1-|\inp{\alpha_0}{\alpha_1}|^2) \cdot d_b $ for the dual for $p$. This value then upper bounds $p$. The dual SDP
for $b$ is strictly feasible in our case, which means that we can make $d_b$ as close to $b$ as we want. This implies
the theorem.

Let $\delta:=\sqrt{1-|\inp{\alpha_0}{\alpha_1}|^2}$. Then we want to show $p \leq 16 \delta^2 b$.
The dual SDP for $b$ is
\begin{equation}\label{eq:dualsingle}
\begin{array}{ll}
 \mbox{minimize} & \Tr[X_b]\\
 \mbox{subject to} & X_b \succeq 0, z_b\geq 0,\\
 & X_b \succeq \frac{1}{2}\left( (1+\eps z_b)\beta_0 + (1-(1-\eps)z_b)\beta_1 \right)=:X_1, \\
 & X_b \succeq \frac{1}{2}\left( (1+\eps z_b)\beta_1 + (1-(1-\eps)z_b)\beta_0 \right)=:X_2.\\
\end{array}
\end{equation}
This SDP is strictly feasible, for example, $z_b=\frac{1}{2}, X_b=2 I$ is a strictly feasible solution.  Hence by
strong duality its optimal value is exactly $b$.

The dual SDP for $p$ is
\begin{equation}\label{eq:dualdouble}
\begin{array}{ll}
\mbox{minimize} & \Tr[X]\\
\mbox{subject to} & X \succeq 0, z \geq 0, \\
            & X \succeq \frac{1}{4}\left(\left\{ (1+\frac{\eps}{2} z)\alpha_0+(1-(1-\frac{\eps}{2})z)\alpha_1\right\}
            \otimes\beta_0+(1-(1-\frac{\eps}{2})z)(\alpha_0 + \alpha_1)\otimes\beta_1 \right)=:X'_1, \\
            & X \succeq \frac{1}{4}\left(\left\{ (1+\frac{\eps}{2} z)\alpha_0+(1-(1-\frac{\eps}{2})z)\alpha_1\right\}
            \otimes\beta_1+(1-(1-\frac{\eps}{2})z)(\alpha_0 + \alpha_1)\otimes\beta_0 \right)=:X'_2,\\
            & X \succeq \frac{1}{4}\left(\left\{ (1+\frac{\eps}{2} z)\alpha_1+(1-(1-\frac{\eps}{2})z)\alpha_0\right\}
            \otimes\beta_0+(1-(1-\frac{\eps}{2})z)(\alpha_0 + \alpha_1)\otimes\beta_1 \right)=:X'_3,\\
            & X \succeq \frac{1}{4}\left(\left\{ (1+\frac{\eps}{2} z)\alpha_1+(1-(1-\frac{\eps}{2})z)\alpha_0\right\}
            \otimes\beta_1+(1-(1-\frac{\eps}{2})z)(\alpha_0 + \alpha_1)\otimes\beta_0 \right)=:X'_4.\\
\end{array}
\end{equation}
For what follows we need to define the positive part of a Hermitian matrix. Any Hermitian matrix $A$ can be written
uniquely as $A=A^+-A^-$, where $A^+,A^-$ are positive semidefinite ($A^+,A^- \succeq 0$) and have orthogonal support.
Then define $\Pos(A)=A^+$. We need the following simple properties:
\begin{claim}\label{claim:pos}
\begin{enumerate}
\item If $A \preceq B$ then $A \preceq \Pos(B)$.
 \item If $A \succeq 0$ then $\Pos(A \otimes B)=A \otimes \Pos(B)$.
 \item If $A \preceq B$ then $\Tr[\Pos(A)] \leq \Tr[\Pos(B)]$.
\end{enumerate}
\end{claim}
Note that it is {\em not} true that if $A \preceq B$ then $\Pos(A) \preceq \Pos(B)$.

\begin{proof}
The first part follows from $B \preceq \Pos(B)$. The second part can be seen by diagonalizing the matrices (note that
the non-zero eigenvalues of $\Pos(B)$ are exactly the positive eigenvalues of $B$). The third part can be seen for
instance by using majorization (see e.g. \cite{Bhatia:97a}). 
If $A \preceq B$, then the vector of eigenvalues of $A$ is
submajorized by the vector of eigenvalues of $B$ (\cite{Bhatia:97a}, Eq.~(II.16), Ky Fan Maximum Principle).
This means that if we
order the eigenvalues of $A$ (resp.~$B$) as  $\lambda_1 \geq \lambda_2 \geq \ldots $ (resp.~$\mu_1 \geq \mu_2 \geq
\ldots$) then for all $k \geq 1$, $\sum_{i=1}^k \lambda_i \leq \sum_{i=1}^k \mu_i$. Together with the fact that the
trace of $\Pos(A)$ is the sum of the positive eigenvalues of $A$, the property follows.
\end{proof}

 We also need the following technical claim, which we
will prove afterwards:

\begin{claim}\label{claim:grr}
Let $0 \leq \eps <1/2$ and $\sigma_0,\sigma_1,\rho_0,\rho_1$ be density matrices, where $\rho_0$ and $\rho_1$  are
$2$-dimensional of rank $1$ (i.e., pure states). Denote by $\rho^\perp_1=I-\rho_1$ the rank $1$ density matrix whose
support is orthogonal to that of $\rho_1$. Then for all $z_b \geq 0$ there exists $ z =z(\eps,z_b)\geq 0$ such that
\begin{align*}
 4 \delta^2 \rho^\perp_1 \otimes & \frac{1}{2}\left\{ (1+\eps z_b) \sigma_0  + (1-(1-\eps)z_b)
\sigma_1 \right\}\\
 & \succeq \frac{1}{4}\left(\left\{(1+\frac{\eps}{2} z) \rho_0 + (1-(1-\frac{\eps}{2})z) \rho_1 \right\} \otimes \sigma_0
+(1-(1-\frac{\eps}{2})z)(\rho_0+ \rho_1) \otimes \sigma_1\right).
\end{align*}
\end{claim}

Fix a dual solution $(X_b,z_b)$ for (\ref{eq:dualsingle}).  Our goal is to find a feasible solution $(X,z)$ to
(\ref{eq:dualdouble}) such that $\Tr[X] \leq 16 \delta^2 \Tr[X_b]$. Since $\ket{\alpha_0}$ and $\ket{\alpha_1}$ are
pure states, we can assume without loss of generality that they are in a two dimensional space, and therefore we can
apply Claim \ref{claim:grr} with $\rho_0=\alpha_0$, $\rho_1=\alpha_1$, $\sigma_0=\beta_0$ and $\sigma_1=\beta_1$. Let
$$
Y_1= 4 \delta^2 \alpha^\perp_1 \otimes \frac{1}{2}\left\{ (1+\eps z_b) \beta_0 + (1-(1-\eps)z_b) \beta_1 \right\}=4 \delta^2 \alpha^\perp_1 \otimes X_1.
$$
Claim~\ref{claim:grr} gives a $z=z(\eps,z_b)$ such that $Y_1 \succeq X'_1$
(see (\ref{eq:dualdouble}) for the definition of $X'_1$). Note that because
$\alpha^\perp_1 \succeq 0$ we can use Claim \ref{claim:pos}.2:
$$
\Pos(Y_1)=4 \delta^2 \alpha^\perp_1 \otimes \Pos\frac{1}{2}\left\{ (1+\eps z_b) \beta_0 + (1-(1-\eps)z_b) \beta_1
\right\}=4 \delta^2 \alpha^\perp_1 \otimes \Pos(X_1).
$$
Because $\alpha^\perp_1\succeq 0$, $\Tr[\Pos (Y_1)]=4 \delta^2 \Tr[\Pos (X_1)]$. Moreover,  
$X_1 \preceq X_b$ by definition (see (\ref{eq:dualsingle})) and $X_b=\Pos(X_b)$, hence $\Tr[\Pos (Y_1)]\leq
4 \delta^2 \Tr[\Pos (X_b)] =4 \delta^2 \Tr[X_b]$ (using Claim \ref{claim:pos}.3).

However, $\Pos (Y_1)$ is not a solution of the dual SDP in (\ref{eq:dualdouble}) because it need not satisfy the last
three inequalities. We construct three more matrices $Y_2$, $Y_3$ and $Y_4$ such that $Y_i \succeq X'_i$ for {\em the
same} $z$ as before. For this we apply Claim \ref{claim:grr} three more times (for $Y_2=4 \delta^2 \alpha^\perp_1
\otimes X_2$ with $(\rho_0, \rho_1, \sigma_0, \sigma_1)=(\alpha_0,\alpha_1 , \beta_1 , \beta_0)$, for $Y_3=4 \delta^2
\alpha^\perp_0 \otimes X_1$ with $(\rho_0, \rho_1, \sigma_0, \sigma_1)=(\alpha_1,\alpha_0 , \beta_0 , \beta_1)$ and for
$Y_4=4 \delta^2 \alpha^\perp_0 \otimes X_2$ with $(\rho_0, \rho_1, \sigma_0, \sigma_1)=(\alpha_1,\alpha_0 , \beta_1 ,
\beta_0)$). Because $z$ depends only on $z_b$ and $\eps$, which are the same in all four applications, 
we obtain each time the same $z$. Now define $X=\sum_{i=1}^4 \Pos(Y_i)$. Clearly $(X,z)$ is a feasible
solution to the SDP (\ref{eq:dualdouble}) since $X \succeq 0$  by definition and $X \succeq \Pos(Y_i) \succeq X_i'$ for
$i=1\ldots 4$ (using Claim \ref{claim:pos}.1). But $\Tr[X] = \sum_{i=1}^4 \Tr[\Pos(Y_i)] \leq 16 \delta^2 \Tr[X_b]$.
As $\Tr[X]$ is an upper bound on $p$, and $\Tr[X_b]$ can be made arbitrarily close to $b$, this implies the theorem.
\end{proof}

\begin{proof}[ of Claim \ref{claim:grr}]
Because $\sigma_0$ and $\sigma_1$ are positive semidefinite, it suffices to
find a $z \geq 0$ for which the equations
\begin{equation}\label{eq:first}
 4 \delta^2 \rho^\perp_1  \frac{1}{2} (1+\eps z_b) \succeq \frac{1}{4}\left\{(1+\frac{\eps}{2} z) \rho_0
 + (1-(1-\frac{\eps}{2})z) \rho_1 \right\}
 \end{equation}
 and
\begin{equation}\label{eq:second}
 4 \delta^2 \rho^\perp_1  \frac{1}{2} (1-(1-\eps)z_b) \succeq
 \frac{1}{4}(1-(1-\frac{\eps}{2})z)(\rho_0+ \rho_1)
\end{equation}
are true.

Let $\ket{\rho_0}, \ket{\rho_1}$ and $\ket{\rho_1^\perp}$ be pure states whose density matrices are $\rho_0,\rho_1$ and
$\rho^\perp_1$. We choose their global phase such that $\ket{\rho_0}=\sqrt{1-\delta^2}\ket{\rho_1}+\delta
\ket{\rho_1^\perp}$. Then, in the basis given by $\ket{\rho_1}$, $\ket{\rho_1^\perp}$, Eqs.~(\ref{eq:first})
and~(\ref{eq:second}) become
\begin{align}\label{eq:firstprime}
\left(\begin{array}{cc}
-(1+\frac{\eps}{2} z)(1-\delta^2)-(1-(1-\frac{\eps}{2})z) & -\delta \sqrt{1-\delta^2}(1+\frac{\eps}{2} z)\\
-\delta \sqrt{1-\delta^2}(1+\frac{\eps}{2} z)& 8\delta^2(1+\eps z_b)-\delta^2(1+\frac{\eps}{2} z)
\end{array} \right)&=\nonumber\\
 \left(\begin{array}{cc}
 z(1-\eps+\delta^2 \frac{\eps}{2})+\delta^2-2 &-\delta \sqrt{1-\delta^2}(1+\frac{\eps}{2} z)\\
-\delta \sqrt{1-\delta^2}(1+\frac{\eps}{2} z) & \delta^2 (7+8 \eps z_b-\frac{\eps}{2}z)
\end{array} \right) &\succeq 0
\end{align}
and
\begin{align}\label{eq:secondprime}
\left(\begin{array}{cc}
 -(1-(1-\frac{\eps}{2})z)(2-\delta^2) & -\delta \sqrt{1-\delta^2}(1-(1-\frac{\eps}{2})z)\\
-\delta \sqrt{1-\delta^2}(1-(1-\frac{\eps}{2})z) & 8 \delta^2(1-(1-\eps)z_b)-\delta^2(1-(1-\frac{\eps}{2})z)
\end{array} \right)&=\nonumber \\
\left(\begin{array}{cc}
 ((1-\frac{\eps}{2})z-1)(2-\delta^2) & \delta \sqrt{1-\delta^2}((1-\frac{\eps}{2})z-1)\\
 \delta \sqrt{1-\delta^2}((1-\frac{\eps}{2})z-1) & \delta^2(7-8(1-\eps)z_b+(1-\frac{\eps}{2})z)
\end{array} \right)&\succeq 0
\end{align}
To show that a $2\times 2$ Hermitian matrix is positive semidefinite it suffices to show that both its determinant and
at least one of its diagonal entries are positive. We choose
$$
z=16 \frac{1-\eps}{1-\eps/2}z_b+\frac{4}{1-\eps}.
$$
Since $z \geq 4$, the upper diagonal entries of the matrices in Eqs.~(\ref{eq:firstprime}) and
(\ref{eq:secondprime}) are positive. Moreover, if $\delta=0$ these matrices are trivially positive.
If $\delta>0$ then we can cancel $\delta^2 >0$ from both terms that appear in their determinants.
Hence, for Eqs.~(\ref{eq:firstprime}) and~(\ref{eq:secondprime}) to be true it suffices to show
\begin{align}\label{eq:firstpp}
 &\big(z(1-\eps)-2\big)(7+8 \eps z_b-\frac{\eps}{2}z) -  (1+\frac{\eps}{2} z)^2 > 0
\end{align}
and
\begin{align}\label{eq:secondpp}
 &(2-\delta^2)((1-\frac{\eps}{2})z-1)(7-8(1-\eps)z_b+(1-\frac{\eps}{2})z) - (1-\delta^2)((1-\frac{\eps}{2})z-1)^2> 0.
\end{align}
To derive Eq.~(\ref{eq:firstpp}) we have replaced the term $ z(1-\eps+\delta^2 \frac{\eps}{2})+\delta^2-2$ by the
smaller positive term $z(1-\eps)-2$, which is legal because this equation is only true if $7+8 \eps
z_b-\frac{\eps}{2}z>0$. Using $(2-\delta^2)/(1-\delta^2)\geq 2$ and $(1-\frac{\eps}{2})z-1>0$, Eq.~(\ref{eq:secondpp})
is implied by
\begin{align*}
 2(7-8(1-\eps)z_b+(1-\frac{\eps}{2})z) & > (1-\frac{\eps}{2})z-1
 \end{align*}
 which is equivalent to
 \begin{align*}
  z & > 16 z_b
\frac{1-\eps}{1-\frac{\eps}{2}}-\frac{15}{1-\frac{\eps}{2}}.
\end{align*}
This inequality is true for our choice of $z$. It remains to show that our $z$ satisfies Eq.~(\ref{eq:firstpp}).
Substituting for $z$ we see that the quadratic term in $z_b$ cancels and we obtain
$$
\Big(17 -\frac{4}{(1-\eps)^2}\Big) +16 z_b \Big(\frac{7}{1-\frac{\eps}{2}}-17 \eps \Big)   > 0.
$$
This linear inequality is satisfied (for $z_b \geq 0$) because both its constant coefficient and the coefficient of $z_b$
are positive for $0\leq \eps < \frac{1}{2}$.
\end{proof}

Using this result, we can as a corollary also prove a second, ``asymmetric" direct product theorem
when $\alpha_0$, $\alpha_1$ and $\beta_0$, $\beta_1$ are all mixed states:

\begin{corollary}\label{tensorcorollary}
Let $0 \leq \eps < \frac{1}{2}$ and $\alpha_0,\alpha_1,\beta_0,\beta_1$ be density matrices. Let
$a=\|\alpha_0-\alpha_1\|_{tr}$, $b=D_{\eps}(\beta_0,\beta_1)$, and
$p=D_{\eps/2}(\alpha_0\otimes\beta_0,\alpha_0\otimes\beta_1,\alpha_1\otimes\beta_0,\alpha_1\otimes\beta_1)$. Then
$p\leq 32 \, a \cdot b$.
\end{corollary}

\begin{proof}
The idea is to work with purifications of $\alpha_0$ and $\alpha_1$.
By Uhlmann's theorem~\cite[p.410]{nielsen&chuang:qc} there exist purifications
$\ket{\tilde{\alpha}_0}$ and $\ket{\tilde{\alpha}_1}$ that preserve the fidelity, i.e.,
$F(\alpha_0,\alpha_1)=F(\ket{\tilde{\alpha}_0},\ket{\tilde{\alpha}_1})=|\inp{\tilde{\alpha}_0}{\tilde{\alpha}_1}|$. Using
known properties of the fidelity~\cite[Section~9.2.3]{nielsen&chuang:qc}, we have
\[ F(\alpha_0,\alpha_1) \geq 1-\|\alpha_0-\alpha_1\|_{tr}=1-a. \]
This implies  $1- |\inp{\tilde \alpha_0}{\tilde \alpha_1}|^2 \leq 2a$. Let $\tilde \alpha_i =
\ketbra{\tilde{\alpha}_i}{\tilde{\alpha}_i}$. Then,
\[p=D_{\eps/2}(\alpha_0\otimes\beta_0,\alpha_0\otimes\beta_1,\alpha_1\otimes\beta_0,\alpha_1\otimes\beta_1) \leq
D_{\eps/2}(\tilde \alpha_0\otimes\beta_0,\tilde \alpha_0\otimes\beta_1,\tilde \alpha_1\otimes\beta_0,\tilde
\alpha_1\otimes\beta_1)
\]
because one can obtain $\alpha_0,\alpha_1$ by tracing out the purification degrees of freedom of
$\tilde{\alpha}_0,\tilde{\alpha}_1$. Theorem~\ref{lemma:tensor} now gives $p \leq 16  (1- |\inp{\tilde \alpha_0}{\tilde
\alpha_1}|^2) \cdot b \leq 32 \, a \cdot b$.
\end{proof}

\section{Shared randomness can be exponentially stronger than quantum communication}\label{secappl1}

\subsection{The problem}

In this section we analyze the following communication problem $P_1$ in the SMP model:
\begin{quote}
{\bf Alice's input:} strings $x,s\in\01^n$, with Hamming weight $|s|=n/2$\\
{\bf Bob's input:} a string $y\in\01^n$\\
{\bf Goal:} the referee should output $(i,x_i,y_i)$ for some $i$ such that $s_i=1$
\end{quote}
We allow the referee some small constant error probability $\eps<1/8$.
In the next two subsections we show that this problem is easy if we have
classical communication and shared randomness, and hard if we have quantum
communication without shared randomness.  More precisely, we will prove:

\begin{theorem}
For the relational problem $P_1$ defined above we have
$$
R^{\parallel,pub}_\eps(P_1)\leq O(\log n)\mbox{ and }Q^{\parallel}_\eps(P_1)\geq\Omega(n^{1/3}).
$$
\end{theorem}

\subsection{Upper bound with classical communication and shared randomness}

Shared randomness gives the parties enough coordination to easily solve this problem. Alice and Bob just send
$(i,x_i,s_i)$ and $(i,y_i)$, respectively, to the referee for $\log(1/\eps)$ public random $i$'s. With probability
$1-\eps$, $s_i=1$ for at least one of those $i$'s and the referee outputs the corresponding $(i,x_i,y_i)$. With
probability $\eps$ he doesn't see an $i$ for which $s_i=1$, in which case he outputs something random. Hence
$R_\eps^{\parallel,pub}(P)\leq O(\log n\log(1/\eps))$.

\subsection{Lower bound for quantum communication with private randomness}

Consider some quantum protocol that solves our problem with error probability $\eps<1/8$, and where the messages that
Alice and Bob send to the referee are at most $q$ qubits long.  Our goal is to show $q\geq\Omega(n^{1/3})$.

First consider the mixed state message $\beta_y$ that Bob sends given input $y$.
For $i\in[n]$, let
$$ \beta_{i0}= \frac{1}{2^{n-1}} \sum_{y:y_i=0} \beta_y$$
be the uniform mixture of all $\beta_y$ with $y_i=0$ and define $\beta_{i1}$ similarly.
Let $b_i=D_{4\eps}(\beta_{i0},\beta_{i1})$.
Then by the random access code argument (Lemma~\ref{lem:RAC}) we have
$$
\sum_{i=1}^nb_i(1-H(4\eps))\leq q.
$$
By Markov's inequality, there is a set $S$ of $n/2$ $i$'s such that $b_i\leq 2q/n(1-H(4\eps)) \le O(q/n)$
for all $i\in S$. We now
fix Alice's input $s$ to be the $n$-bit string with support corresponding to $S$.

We now analyze Alice's message.
Let $\alpha_x$ be the mixed state she sends given input $x$ and our fixed $s$.
Define $\alpha_{i0}$
as the uniform mixture of all $\alpha_x$ with $x_i=0$, similarly define $\alpha_{i1}$, and
$a_i=\norm{\alpha_{i0}-\alpha_{i1}}_{tr}$. The optimal probability with which we can distinguish $\alpha_{i0}$ from
$\alpha_{i1}$ is $\frac{1}{2}+\frac{a_i}{2}$. The random access code argument
gives
$$
\sum_{i=1}^n a_i^2\leq O(q).
$$

Now we look at the protocol's behavior.
Let $X=X_1\ldots X_n$ and $Y=Y_1\ldots Y_n$ be uniformly distributed random variables
giving Alice's first and Bob's only input, and $I$, $B_1$, $B_2$
be the random variables describing the referee's output.
We call an index $i\in[n]$ \emph{good}, if the protocol is correct
with high probability when it outputs $(i,*,*)$:
\begin{center}
$i$ is good iff $i\in S$ and $\Pr[B_1=X_i,B_2=Y_i\mid I=i]\geq 1-2\eps$.
\end{center}
The index is called \emph{bad} otherwise.
Define $p_i=\Pr[I=i]$ to be the probability that the referee outputs something of
the form $(i,*,*)$. Because the protocol is correct with probability at least $1-\eps$,
a Markov argument shows that the good indices must together have most of the probability:
$$
1-\eps\leq \sum_{{\rm good}\; i}p_i + \sum_{{\rm bad}\; i} (1-2\eps)p_i=1-2\eps+2\eps\sum_{{\rm good}\; i}p_i,
$$
hence
$$
\frac{1}{2}\leq \sum_{{\rm good}\; i}p_i.
$$
Notice that for each good $i$ we can use the protocol to get a $(p_i,2\eps)$-predictor for $X_iY_i$:
just run the protocol and return `$\dono$' if the protocol's output is not of the form $(i,*,*)$,
and otherwise return the last two bits of the protocol's output.
Therefore Corollary~\ref{tensorcorollary} implies $p_i\leq O(a_i b_i)$.
Also, $b_i\leq O(q/n)$ for all good $i$ so we can bound
$$
\frac{1}{2}\leq \sum_{{\rm good}\; i}p_i \leq \sum_{{\rm good}\; i}O(a_i b_i)\leq O\left(\frac{q}{n}\sum_{i=1}^n a_i\right)\leq
O\left(\frac{q}{n}\sqrt{n\sum_{i=1}^n a_i^2}\right)\leq O\left(\frac{q^{3/2}}{n^{1/2}}\right),
$$
where we applied Cauchy-Schwarz in the fourth step. This implies $q\geq\Omega(n^{1/3})$.

\paragraph{\bf Remark:} The best no-shared-randomness protocol we know for $P_1$ communicates $O(\sqrt{n})$ bits. The idea is to arrange the
$n$-bit inputs in a $\sqrt{n}\times\sqrt{n}$ matrix. Alice picks a random row index in $[\sqrt{n}]$, and then sends
that index and the indexed row of $x$ and of $s$ to the referee. Bob picks a random column index in $[\sqrt{n}]$, and
then sends that index and the indexed column of $y$ to the referee. The row and the column intersect in exactly one
(uniformly random) point $i\in[n]$. With probability 1/2, $s_i=1$ and we are done. Repeating this a few times in
parallel reduces the error probability to a small constant. A matching lower bound would follow from the general direct
product theorem $p\leq O(ab)$, for the case of the 2-register identification problem where both sides are allowed to be
mixed.

\section{Entanglement can be exponentially stronger than quantum communication with shared randomness}\label{secappl2}

\subsection{The problem}

For $n$ a power of 2, consider the following relational problem $P_2$,
inspired by a one-way communication problem due to Bar-Yossef et~al.~\cite{bjk:q1way}:
\begin{quote}
{\bf Alice's input:} a perfect matching $M\subset{[n]\choose 2}$ and a string $x\in\01^{n/2}$
containing a bit $x_e$ for each edge $e\in M$\\
{\bf Bob's input:} a string $y\in\01^n$\\
{\bf Goal:} the referee should output $(i,j,x_{(i,j)},y_i\oplus y_j)$ for some edge $(i,j)\in M$
\end{quote}
Below we show that this problem is easy if we have classical communication and prior entanglement,
and hard if we have quantum communication without entanglement:

\begin{theorem}
For the relational problem $P_2$ defined above we have
$$
R^{\parallel,ent}_\eps(P_2)\leq O(\log n)\mbox{ and }Q^{\parallel,pub}_\eps(P_2)\geq\Omega((n/\log n)^{1/3}).
$$
\end{theorem}

\subsection{Upper bound with classical communication and entanglement}\label{secp2upperbound}

The following protocol solves the problem with success probability 1,
using $O(\log n)$ classical bits of communication and $\log n$
EPR-pairs shared between Alice and Bob.  It is a modification of
an unpublished protocol due to Harry Buhrman~\cite{buhrman:matchingcomm}, 
which is in turn based on a one-way protocol from~\cite{bjk:q1way}.
The starting state of Alice and Bob is
$$
\frac{1}{\sqrt{n}}\sum_{i\in\01^{\log n}}\ket{i}\ket{i}.
$$
Bob adds his bits as phases:
$$
\frac{1}{\sqrt{n}}\sum_i \ket{i}(-1)^{y_i}\ket{i}.
$$
Alice measures with the $n/2$ projectors $E_{ij}=\ketbra{i}{i}+\ketbra{j}{j}$ induced by the $n/2$ pairs $(i,j)\in M$.
This gives her a random $(i,j)\in M$ and the resulting joint state of Alice and Bob is
$$
\frac{1}{\sqrt{2}}\left(\ket{i}(-1)^{y_i}\ket{i}+\ket{j}(-1)^{y_j}\ket{j}\right).
$$
Now both players apply a Hadamard transform to each of the $\log n$ qubits of their part of the state, which becomes
(ignoring normalization)
$$
\sum_{k,\ell}\left( (-1)^{y_i+(k+\ell)\cdot i}+(-1)^{y_j+(k+\ell)\cdot j}  \right) \ket{k}\ket{\ell}.
$$
Note that $\ket{k}\ket{\ell}$ has non-zero amplitude iff
$y_i+(k+\ell)\cdot i=y_j+(k+\ell)\cdot j\mod 2$, equivalently
$$
(k+\ell)\cdot (i+j)=y_i\oplus y_j.
$$
Alice and Bob both measure their part of the state in the computational basis, obtaining some $k$ and $\ell$,
respectively, satisfying the above equality. Alice sends $i,j,k$, and $x_{(i,j)}$ to the referee, Bob sends $\ell$; a
total of $O(\log n)$ bits of communication. The referee calculates $y_i\oplus y_j$ from $i,j,k,\ell$ and outputs
$(i,j,x_{(i,j)},y_i\oplus y_j)$ as required.

\subsection{Lower bound for quantum communication without entanglement}

We make use of some ideas from the classical lower bound of Bar-Yossef et~al.~\cite{bjk:q1way}. For $k\in\{0,\ldots,n/2-1\}$,
let $M_k$ denote the matching $\{(i,(i+k-1\mod n/2)+n/2+1\}_{i=1}^{n/2}$. For example,
$M_1=\{(1,n/2+2),(2,n/2+3),(3,n/2+4),\ldots,(n/2-1,n),(n/2,n/2+1)\}$. We will prove our lower bound for the special
case where Alice's matching is one of the $M_k$. Consider a quantum protocol where Alice and Bob share randomness but
no entanglement, each communicates at most $q$ qubits to the referee, and they solve problem $P_2$ with error
probability $\eps<1/16$ for each input. Our goal is to show $q\geq\Omega((n/\log n)^{1/3})$.

We consider the following input distribution.
Let $K$ be a uniformly random number between 0 and $n/2-1$, $M_K$ be Alice's first input,
and $X\in\01^{n/2}$ and $Y\in\01^n$ be uniformly distributed random variables
for Alice's second and Bob's only input.
Since the protocol has error at most $\eps$ for all inputs, we can
(and will) fix a value for the shared randomness such that the resulting
protocol has average error at most $\eps$ under the above input distribution.

Let $\alpha_{kx}$ be Alice's message on input $M_k,x$. For edge $e=(i,j)\in M_k$, define $\alpha_{ke0}$
as the uniform mixture of all $\alpha_{kx}$ with $x_e=0$, similarly define $\alpha_{ke1}$, and
$a_{ke}=\norm{\alpha_{ke0}-\alpha_{ke1}}_{tr}$. The optimal probability with which we can distinguish $\alpha_{ke0}$ from
$\alpha_{ke1}$ is $1/2+a_{ke}/2$. Hence for every $k$, the random access code argument (Lemma~\ref{lem:RAC}) gives
$$
\sum_{e\in M_k} a_{ke}^2\leq O(q).
$$
Let $\beta_y$ be Bob's message on input $y$.
For any $e=(i,j)$ (not necessarily part of any matching),
define $\beta_{e0}$ as the uniform mixture over all $\beta_y$
with $y_i \oplus y_j = 0$ and similarly define $\beta_{e1}$.
Let $b_e = D_{8\eps}(\beta_{e0},\beta_{e1})$.
We now prove two claims upper bounding sums of these $b_e$.

\begin{claim}\label{claim:forest}
For any forest (i.e., acyclic graph) $F$ on $[n]$ we have $\displaystyle\sum_{e\in F}b_e\leq O(q)$.
\end{claim}

\begin{proof}
Denote by $|F|$ the number of edges in $F$. For every $e=(i,j)\in F$ we can obtain a $(b_e,8\eps)$-predictor for the
bit $Y_i\oplus Y_j$ given the $q$-qubit state $\beta_Y$. Intuitively, since $F$ is a forest, these $|F|$ bits are
independent and therefore represent $|F|$ bits of information. To make this formal, define for each $w \in \01^{|F|}$
the set
$$
T_w = \{y \in \01^n \mid \forall e=(i,j) \in F, y_i \oplus y_j = w_e\}.
$$
Since $F$ is a forest, $\{T_w\}_{w \in \01^{|F|}}$ is a partition
of $\01^n$ into $2^{|F|}$ sets of size $2^{n - |F|}$ each.

For any bit string $w \in \01^{|F|}$ we define $\xi_w$ as
the uniform mixture of $\beta_y$ over all $y \in T_w$. For each $e \in F$, define
$\xi_{e0}$ as the uniform mixture of $\xi_w$ over all $w$ with $w_e=0$ and similarly
define $\xi_{e1}$. Then, it is easy to see that $\xi_{e0}=\beta_{e0}$ and
$\xi_{e1}=\beta_{e1}$. Hence, $D_{8 \eps}(\xi_{e0},\xi_{e1}) = b_e$ and
by applying the random access code argument to the encoding of $w$ as the $q$-qubit state $\xi_w$, we get
$$
\sum_{e\in F}b_e(1-H(8\eps))\leq q.
$$
\end{proof}

\begin{claim}\label{claim:edges}
$\displaystyle\sum_{k=0}^{n/2-1}\sum_{e\in M_k} b_e^2\leq O(q^2\log n)$.
\end{claim}

\begin{proof}
By construction all our $M_k$'s are disjoint, hence the set
$M=\cup_{k}M_{k}$ contains each edge in the above sum exactly once.
Making some bijection between edges in $M$ and numbers $\ell\in[|M|]$,
we order the $b_e$ in non-increasing order as
$$
b_1\geq b_2\geq\cdots\geq b_{|M|}.
$$
Now consider the graph consisting of the first $\ell$ edges in this ordering. This graph must contain at least
$\sqrt{2\ell}$ non-isolated vertices, since $v$ vertices give only ${v\choose 2}\leq v^2/2$ distinct edges. Let $F$ be
a forest consisting of a spanning tree for each connected component of this graph. This $F$ has at least
$\sqrt{2\ell}/2=\sqrt{\ell/2}$ edges, and for each of those edges $e$ we have $b_e\geq b_\ell$. Now we can use Claim
\ref{claim:forest}:
$$
\sqrt{\frac{\ell}{2}}\cdot b_\ell \leq \sum_{e\in F} b_e\leq O(q).
$$
Hence for all $\ell\leq|M|$ we have
$$
b_\ell\leq O(q/\sqrt{\ell}).
$$
Summing over all $\ell$ gives
$$
\sum_{e\in M} b_e^2=\sum_{\ell=1}^{|M|} b_\ell^2 \leq \sum_{\ell=1}^{n^2/4} O(q^2/\ell) \leq O(q^2\log n).
$$
\end{proof}

Since the protocol has average error at most $\eps$, by Markov's inequality there is a set $\cal M$ of at least $n/4$
of our matchings $M_k$ such that the protocol has error at most $2\eps$ for that $M_k$ and uniformly random $X$ and
$Y$. Since $\cal M$ contains at least $n/4$ elements, Claim \ref{claim:edges} implies there is a matching $M_k\in{\cal
M}$ such that
$$
\sum_{e\in M_k} b_e^2\leq O\left(\frac{q^2\log n}{n}\right).
$$
We now fix this matching on Alice's side.
Let $I,J,B_1,B_2$ be the random variables giving the referee's output.
Suppose we run the protocol with $M_k$, and uniformly random $x$ and $y$
as input. We call an edge $(i,j)$ \emph{good}, if the protocol is correct
with high probability when it outputs $(i,j,*,*)$:
\begin{center}
$e=(i,j)$ is good iff $e\in M_k$ and $\Pr[B_1=X_e,B_2=Y_i\oplus Y_j\mid I=i,J=j]\geq 1-4\eps$.
\end{center}
The edge is called \emph{bad} otherwise.
Let $p_e=\Pr[I=i,J=j]$ be the probability that the protocol outputs edge $e$.
Since $M_k\in{\cal M}$, the success probability (averaged over $x$ and $y$)
is at least $1-2\eps$, so by a Markov argument, the good edges must have most of the probability:
$$
1-2\eps \leq \sum_{{\rm good}\; e}p_e + \sum_{{\rm bad}\; e} p_e(1-4\eps)
=1-4\eps + 4\eps\sum_{{\rm good}\; e}p_e,
$$
hence
$$
\frac{1}{2}\leq\sum_{{\rm good}\; e}p_e.
$$
For every good edge $e$, we can construct a $(p_e,4\eps)$-predictor for $(X_e,Y_i \oplus Y_j)$.
Hence, by Corollary~\ref{tensorcorollary}, $p_e\leq O(a_{ke} b_e)$.
Using Cauchy-Schwarz:
$$
\frac{1}{2}
\leq\sum_{{\rm good}\; e}p_e
\leq \sum_{{\rm good}\; e}O(a_{ke} b_e)
\leq O\left(\sqrt{ \sum_{{\rm good}\; e}a_{ke}^2 \cdot \sum_{{\rm good}\; e}b_e^2 }\right)
\leq O\left( \sqrt{\frac{q^3\log n}{n}}\right).
$$
This implies the promised lower bound $q\geq\Omega((n/\log n)^{1/3})$.

\paragraph{\bf Remark:} Our bound is tight up to $\log n$ factors. To see this, we briefly sketch a protocol which uses $O(n^{1/3}\log n)$
qubits of communication: Alice and Bob use their shared randomness to fix a subset $S \subset [n]$ of size $n^{2/3}$.
With high probability the number of edges from $M$ contained in $S\times S$ is roughly $n^{1/3}$. For each of the edges
$(i,j)\in M\cap S\times S$, Alice sends $(i,j,x_{(i,j)})$ to the referee, which is $O(n^{1/3} \log n)$ bits of
communication. Bob prepares $n^{1/3}$ copies of the state
\begin{equation}\label{eq:bobstate}
\frac{1}{\sqrt{|S|}} \sum_{i \in S} (-1)^{y_i} \ket{i}
\end{equation}
and sends them to the referee. This gives a total of $O(n^{1/3} \log n)$ qubits of communication. On each of the
copies, the referee measures with the projectors $E_{ij}=\ketbra{i}{i}+\ketbra{j}{j}$ induced by the edges in $S$ that
Alice has sent, completed by $E_{garbage}=I-\sum E_{ij}$. Given the state in Eq.~(\ref{eq:bobstate}), the probability to
not measure ``garbage" is roughly $n^{-1/3}$. This means that with some constant probability the referee will measure
one of the edges $E_{ij}$ on one of the states Bob sent. This state then collapses to
$\frac{1}{\sqrt{2}}((-1)^{y_i}\ket{i}+(-1)^{y_j}\ket{j})$, and a measurement in the basis $\ket{i} \pm \ket{j}$ gives
$y_i \oplus y_j$.

\section{Conclusion and future work}

We studied the bounded-error quantum state identification problem and proved a direct product theorem for two
independent instances of this problem (one involving pure states) using SDP duality. We applied our direct product
theorem to obtain two exponential separations in the simultaneous message passing model of communication complexity.
These two separations nicely complement each other: the first shows that shared randomness is much more powerful
than private randomness, the second shows that prior entanglement is much more powerful than shared randomness. Moreover,
both separations are shown in the strongest possible sense: the stronger model is restricted to classical communication
while the weaker model is allowed quantum communication.

We identify some interesting problems left open by our work.
First, for the bounded-error quantum state identification problem,
prove the direct product theorem $p\leq O(ab)$ in the general case where
both sides have mixed states instead of one side pure and one side mixed.
That result would lift, for instance, our quantum communication lower bound
for the problem $P_1$ to the optimal $\Omega(\sqrt{n})$.
Second, show similar communication complexity separations for
decision problems (Boolean functions, possibly with a promise on the input)
instead of for relational problems.
Finally, we hope our direct product theorem will be useful
for other applications as well.

\subsection*{Acknowledgments}
We thank Harry Buhrman for permission to include his protocol, which we eventually modified to the protocol of
Section~\ref{secp2upperbound}. DG is grateful to Richard Cleve for helpful discussions.

\bibliographystyle{plain}

\end{document}